%% file: article.tex
\documentclass[onefignum,onetabnum]{siamonline250211}
\usepackage{bm}

\input{ex_shared}

\ifpdf
\hypersetup{
  pdftitle={An Example Article},
  pdfauthor={Y. Abdullah, V. Hart, and M. Das}
}
\fi

\begin{document}

\maketitle

\begin{abstract}

We investigate a conductance‑based neuron model to explore how voltage‑gated ion channel isoforms influence action‑potential generation. The model combines a six‑state Markov representation of NaV channels with a first‑order  K\textsubscript{V}3.1 model, allowing us to vary maximal sodium and potassium conductances and compare nine NaV isoforms. Using bifurcation theory and local stability analysis, we map regions of stable limit cycles and visualize excitability landscapes via heatmap‑based diagrams. These analyses show that isoforms NaV1.3, NaV1.4 and NaV1.6 support broad excitable regimes, while isoforms NaV1.7 and NaV1.9 exhibit minimal oscillatory behavior. Our findings provide insights into the role of channel heterogeneity in neuronal dynamics and may help to guide the design of synthetic excitable systems by narrowing the parameter space needed for robust action‑potential trains.

\end{abstract}

\begin{keywords}
stability analysis, computational neuroscience
\end{keywords}

\begin{MSCcodes}
68Q25, 68R10, 68U05
\end{MSCcodes}

\section{Introduction}
An action potential is a rapid, transient change in a neuron's membrane electrical potential that serves as the fundamental mechanism for transmitting electrical signals in neurons and other excitable cells. This excitation is essential for the proper functioning of neurons in the Central Nervous System (CNS), Peripheral Nervous System (PNS) and for other excitable cells, such as cardiac muscle and some endocrine cells \cite{princ-neur-sci, stojilkovic_ion_2010, wei_physiology_2025}. Action potential generation is induced by the relative concentrations of ions inside and outside the cell, as well as the membrane’s selective permeability to those ions \cite{Grider_2023}. It is the primary mechanism by which neurons process information and communicate over long distances by propagating action potentials along their axons, which are thin, elongated projections specialized for electrical conduction \cite{princ-neur-sci}.

The Hodgkin–Huxley (HH) model, introduced in 1952, has long served as a foundational framework in electrophysiology for describing the generation and propagation of action potentials in neurons \cite{hodgkin1952quantitative}. Although it successfully reproduces action potential waveforms, particularly in the giant squid axon, it remains fundamentally phenomenological rather than mechanistic. The HH formalism has been shown to exhibit limitations in accurately capturing the detailed electrophysiological behavior of ion channels \cite{Patlak1991,Strassberg1993,Meunier2002,Maurice2004}. Specifically, it assumes a fixed kinetic structure with a small number of independent gates and does not correctly account for Na$^{+}$ channel inactivation kinetics \cite{Bezanilla_Armstrong_1977, Armstrong_Bezanilla_1977}.  For these reasons we need an alternative framework to mathematically model and accurately analyze findings that have been made over the past few decades about different channel isoforms and how they are finely distributed on different part of neuron morphology to carry out specific functions.

Ion channel isoforms are distinct protein variants that selectively conductive to the same ion but are encoded by different genes. Each isoform has slightly different protein sub-units, and consequently, exhibits unique kinetic properties that tailor it to a specific  function within the nervous system \cite{gustafson1993mutually, southan2016iuphar, Lai_Jan_2006}. In humans, nine Na$^{+}$ channel isoforms (Na\textsubscript{V}1.1–1.9) have been identified, each with distinct kinetics, gating behaviors, and tissue distributions \cite{catterall2005international, goldin1999diversity, Wang_Ou_Wang_2017}. These isoforms demonstrate highly specialized, voltage–dependent gating mechanisms, such that even minor alterations in their kinetics can profoundly affect neuronal excitability and system-level dynamics. Indeed, small genetic mutations in Na\textsubscript{V} channels underlie numerous severe channelopathies, particularly in peripheral nerves \cite{sb2011voltage,Wang_Ou_Wang_2017}.

The expression of voltage‑gated sodium channel isoforms is tightly regulated across tissues, supporting both physiological specialization and disease susceptibility.  In the CNS, Na\textsubscript{V}1.1, Na\textsubscript{V}1.2, Na\textsubscript{V}1.3 and Na\textsubscript{V}1.6 are the predominant isoforms \cite{Vacher_Mohapatra_Trimmer_2008,Wang_Ou_Wang_2017}.  By contrast, Na\textsubscript{V}1.7, Na\textsubscript{V}1.8 and Na\textsubscript{V}1.9 are largely restricted to the PNS \cite{Wang_Ou_Wang_2017}.  Within the CNS, Na\textsubscript{V}1.1 and Na\textsubscript{V}1.2 play complementary roles in inhibitory and excitatory neurons, respectively \cite{Candenas_Seda_Noheda_Buschmann_Cintado_Martin_Pinto_2006}.  In excitable non‑neuronal tissues, Na\textsubscript{V}1.4 is the principal skeletal‑muscle isoform: it drives the action potentials that trigger muscle contraction, and mutations in Na\textsubscript{V}1.4 give rise to hereditary myopathies.  Na\textsubscript{V}1.5 is the primary cardiac isoform, expressed in atrial, Purkinje and ventricular cells, where it mediates the rapid depolarization of the cardiac action potential; pathogenic variants in its gene are strongly associated with life‑threatening arrhythmia \cite{Candenas_Seda_Noheda_Buschmann_Cintado_Martin_Pinto_2006}.

Collectively, this isoform-specific distribution highlights the complexity of sodium channel physiology and highlights the limitations of classical HH-type formulations that treat sodium channel dynamics as homogeneous. Accurately modeling these isoforms is essential not only for advancing our understanding of excitability and signal propagation in different tissues but also for the rational design of therapeutic interventions and for developing synthetic excitable cells. In synthetic neuroscience and bioengineering, incorporating isoform-specific dynamics will be critical for creating biologically realistic neural models and for developing targeted therapies for channelopathies and excitability disorders \cite{Alsaloum_Dib-Hajj_Page_Ruben_Krainer_Waxman_2025}.
    
Markov models offer a more flexible and mechanistic alternative to the HH model and are suitable to model these channel isoforms within a unified model. By allowing for arbitrarily many discrete channel states and voltage-dependent transitions between them, they can capture complex gating behaviors and subtle differences in different neurons due to different ion channel isoforms often observed in mammalian ion channels. 

Despite their versatility, Markov models introduce a high-dimensional parameter space due to their multiple states and voltage-dependent transition rates, making their global dynamical behavior difficult to characterize \cite{marder2011multiple}, and consequently is poorly understood. Understanding how changes in conductivity or transition kinetics affect neuronal excitability remains an open challenge. We address this gap by performing a local stability analysis of a hybrid neuron model composed of a six-state Markov model for voltage-gated sodium channels (VGSCs) and a first-order kinetic model for voltage-gated potassium channels (VGPCs). This minimal model is capable of representing a range of neuronal types by varying channel isoforms and parameter values, reflecting the diversity found across different organisms and brain regions \cite{balbi2017single, rudy2001}.

We perform a dynamical stability analysis of the mathematical model partly developed by Balbi et al \cite{balbi2017single} over an extended region in the parameter space of sodium and potassium conductivity to assess the robustness of neuronal behavior under physiological fluctuations in these two parameters. We use bifurcation theory to chart a region of stable oscillatory activity across the sodium-potassium conductivity parameter space. By validating our results with time-series simulations we show the utility of using dynamical stability analysis approach to scan large regions of the parameter space of similar high-dimensional systems. Furthermore, our results provide a flexible framework for predicting how different channel combinations influence excitability, with potential applications in synthetic biology and computational neuroscience. This is particularly relevant in synthetic biology, where such knowledge can guide the bottom-up design of synthetic neurons that aim to replicate the electrophysiological properties of biological counterparts under diverse and potentially noisy conditions.

\section{Mathematical Model}
\label{sec:math-model}

To model the biophysical dynamics underlying action potential generation, we construct a minimal yet physiologically grounded mathematical model of spherical neuron modeled using a single compartment model. The electro-physiological properties of neuronal membrane are modeled by the equivalent electrical circuit formalism introduced by Hodgkin and Huxley \cite{hodgkin1952quantitative}. The membrane potential of a neuron arises from the selective permeability of the membrane to different ionic species, primarily Na$^+$, K$^+$, and Cl$^-$. This ionic selectivity, in conjunction with concentration gradients maintained by active ion pumps, gives rise to the resting membrane potential.

\begin{figure}[H]
    \centering
    \includegraphics[width=0.6\linewidth]{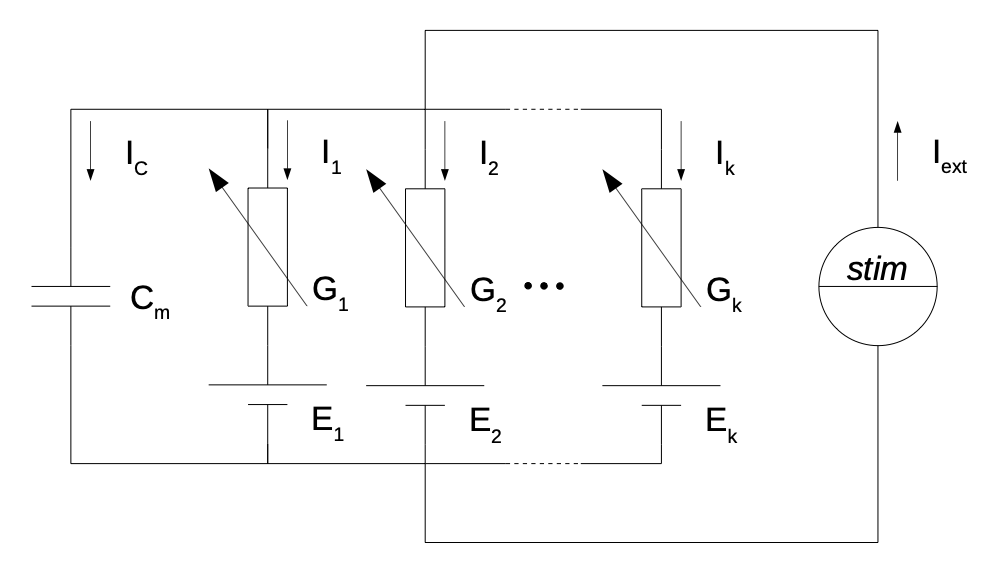}
    \caption{A simple equivalent electrical circuit for a neuronal membrane}
    \label{fig:equiv-circuit}
\end{figure}

A patch of neuronal membrane is modeled as an equivalent electrical circuit (Figure \ref{fig:equiv-circuit}), where ion channels are treated as variable resistances and the membrane’s capacitance represents the lipid bilayer’s ability to store charge. Each ionic current is modeled as a battery-driven resistor, where the battery corresponds to the reversal potential $E_i$ and the resistor's conductance $G_i$ depends on channel gating. In this model, the battery represents the electrochemical driving force for each ion species—the difference between the membrane potential and the ion's equilibrium potential. This reflects the tendency of ions to flow across the membrane due to both concentration and voltage gradients. Finally, an external current is introduced to model the introduction of stimulating signals to the cell through its synapses.

Applying Kirchhoff's current law to this circuit yields the membrane voltage dynamics:
\begin{equation}
\label{eq:kirchhoff}
C_m\frac{dV_m}{dt}+ I_{ion}=I_{ext},
\end{equation}
where $C_m$ is the membrane capacitance, $V_m$ the membrane potential, $I_{ion}$ the total ionic current, and $I_{ext}$ an external stimulus current. The ionic current for each ion species is given by:
\begin{equation}
\label{eq:ohm}
I_{ion} = \sum_i G_i(V_m - E_i).
\end{equation}

To focus on the minimal ionic currents required for action potential generation, we restrict the model to Na$^+$ and K$^+$ channels. These yield two ionic currents, $I_{Na}$ and $I_K$, leading to:
\begin{equation}
\label{eq:model-membrane-voltage}
\frac{dV_m}{dt} = \frac{1}{C_m}(I_{ext} - I_{Na} - I_K).
\end{equation}

To accurately capture the complex gating kinetics of voltage-gated sodium channels, we employ a six-state Markov model originally developed by Balbi et al.~\cite{balbi2017single}. This model allows for transitions among multiple closed, open, and inactivated states and can reproduce the diverse activation, inactivation, and recovery properties observed across all nine human (Na\textsubscript{V}1.1–1.9) isoforms by tuning voltage-dependent transition rates. We limit the use of Markov models to the Balbi et al. Markov model for the sodium channel isoforms, and we use a Hodgkin–Huxley-type kinetic model for the delayed rectifier potassium current using a simpler first-order scheme. This choice leverages the comparatively straightforward activation dynamics of potassium channels and enables faster numerical integration, making the model more computationally efficient without compromising essential biophysical phenomena. Additionally, this choice allows for a relatively simple systematic comparison of the different sodium channel isoforms kinetics.

\begin{figure}[H]
    \centering
    \includegraphics[width=0.8\linewidth]{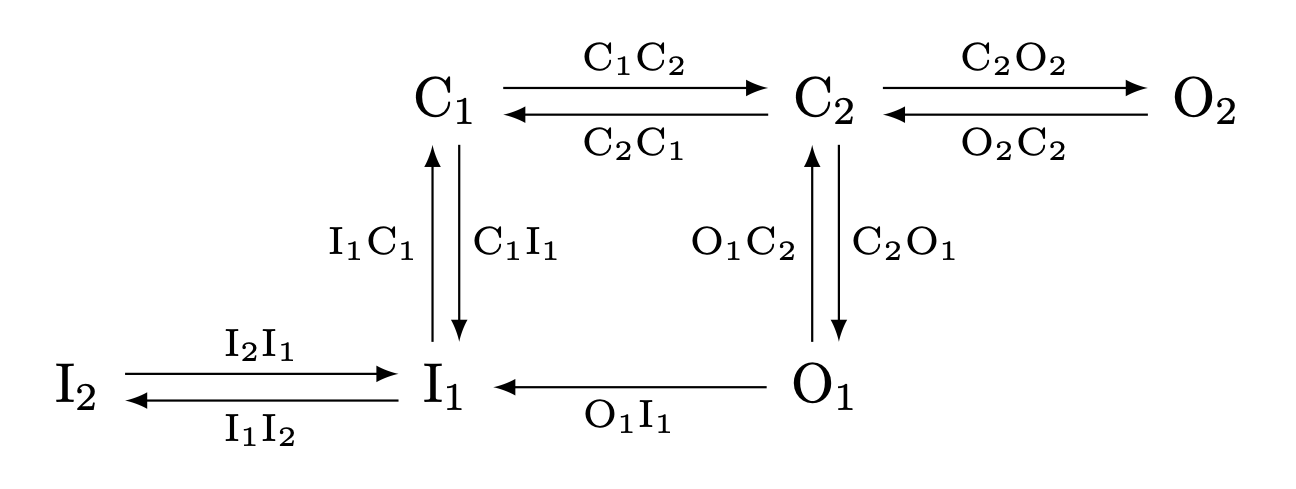}
    \caption{Generalized six-state Markov model for voltage-gated sodium channels. The kinetic scheme includes: Two \textbf{closed} states (C\textsubscript{1}, C\textsubscript{2}) Two \textbf{open} states (O\textsubscript{1}, O\textsubscript{2}) Two \textbf{inactivated} states (I\textsubscript{1}, I\textsubscript{2})}
    \label{fig:sodium-markov-model}
\end{figure}

The six-state Markov model introduced by Balbi et al.~\cite{balbi2017single} was developed by globally fitting voltage-clamp recordings from all nine human Na\textsubscript{V} isoforms to a shared kinetic scheme, enabling a unified representation of sodium channel dynamics across isoforms. Consequently, this Markov model topology does not change from one isoform to another, the only changes are in the transition rate parameters between different states. It consists of two closed states (C\textsubscript{1}, C\textsubscript{2}), two open states (O\textsubscript{1}, O\textsubscript{2}), and two inactivated states (I\textsubscript{1}, I\textsubscript{2}), allowing the model to account for both fast activation and slow inactivation kinetics observed in experimental recordings. Transitions between these states are voltage-dependent and reversible, with transition rates modeled using the sum of two sigmoidal functions, a hyperpolarizing and a depolarizing component: 
\begin{equation}
\label{eq:double-sigmoid}
r_{ij}(V) = B^{\mathrm{hyp}}_{ij} \left[1 + e^{\frac{(V - V^{\mathrm{hyp}}_{ij})}{K^{\mathrm{hyp}}_{ij}}} \right]^{-1} + B^{\mathrm{dep}}_{ij} \left[1 + e^{\frac{(V - V^{\mathrm{dep}}_{ij})}{K^{\mathrm{dep}}_{ij}}} \right]^{-1}
\end{equation}
This double-sigmoid form allows each rate \( r_{ij}(V) \) to flexibly capture both low-voltage activation and high-voltage saturation behavior, improving the fit to experimental data across a wide range of membrane potentials and enabling the unified modeling of diverse Na\textsubscript{V} isoforms. The parameters of this equation are: $B^{\mathrm{hyp}}_{ij}$: The maximal transition rate (in ms\textsuperscript{-1}) for the hyperpolarizing component of the transition from state $j$ to state $i$. It determines the scale of the rate at hyperpolarized membrane potentials. $V^{\mathrm{hyp}}_{ij}$: The half-activation voltage (in mV) of the hyperpolarizing sigmoid. It is the voltage at which the hyperpolarizing term reaches half of its maximum value.  $K^{\mathrm{hyp}}_{ij}$: The slope factor (in mV) controlling the steepness of the hyperpolarizing sigmoid. A smaller value leads to a steeper voltage response. $B^{\mathrm{dep}}_{ij}$: The maximal transition rate (in ms\textsuperscript{-1}) for the depolarizing component. It contributes primarily at more depolarized voltages. $V^{\mathrm{dep}}_{ij}$: The half-activation voltage (in mV) of the depolarizing sigmoid. $K^{\mathrm{dep}}_{ij}$: The slope factor (in mV) for the depolarizing sigmoid. It modulates how sharply the transition rate increases or saturates with depolarization.

These parameters are independently optimized for each transition \( r_{ij}(V) \) using experimental voltage-clamp data for the nine human Na\textsubscript{V} channel isoforms (Na\textsubscript{V}1.1--1.9), allowing the model to capture isoform-specific gating kinetics and voltage dependence. Values of all the parameters can be found in \cite{balbi2017single}.

The dynamics of the voltage-gated sodium channel are governed by a set of six coupled differential equations describing the time evolution of the occupation density of each of the conformational states: two closed states \(C_1\) and \(C_2\), two open states \(O_1\) and \(O_2\), and two inactivated states \(I_1\) and \(I_2\). For brevity we use the vector notation to represent these states: $ \vec{s}(t) = [s_1(t), \dots, s_6(t)]^T = [C_1, C_2, O_1, O_2, I_1, I_2]^T $. In this notation the transition-rate system of linear differential equations can be written as:
\begin{equation}
    \frac{d\vec{s}}{dt} = Q(V)\vec{s}
\end{equation}
where $ Q(V) \in \mathbb{R}^{6 \times 6} $ is the voltage-dependent transition rate matrix, defined such that $ Q_{ij}(V) $ is the rate of transition from state $ j $ to state $ i $, and diagonal entries satisfy
$
    Q_{ii} = -\sum_{j \neq i} Q_{ji}
$
to ensure occupation density conservation:
$
    \sum_{i=1}^6 s_i(t) = 1 \quad \forall t.
$

The macroscopic sodium current is computed by summing the contributions of all open channels, weighted by the driving force and the maximal conductance. In the six-state model, current flows only when the channel is in one of the open states, O\textsubscript{1} or O\textsubscript{2}. Consequently, the total sodium current is given by:
\begin{equation}
I_{\mathrm{Na}} = \bar{g}_{\mathrm{Na}} \left(\mathrm{O}_1 + \mathrm{O}_2\right)(V_m - E_{\mathrm{Na}}),
\end{equation}
where \( \bar{g}_{\mathrm{Na}} \) is the maximal sodium conductance and \( E_{\mathrm{Na}} \) is the sodium reversal potential. This formulation links the Markov state dynamics to the overall electrophysiological behavior of the neuron by capturing the voltage- and time-dependent transition rate of channel opening.

To describe the outward potassium current, we use a single potassium channel isoform K\textsubscript{V}3.1. This voltage gated potassium is often found in the mammalian nervous system in cells that are required to fire often and in high frequency \cite{Rudy2001Kv3}. This choice is motivated by our specific application domain, since we are looking for regions in the parameters space at which a neuron can support action potential trains, and this needs the neurons to be highly active. Furthermore, specific channel isoform is relatively simple to model, and we adopt a classical first-order kinetic model based on the Hodgkin–Huxley formalism. This modeling choice reflects the comparatively simple fast activation-deactivation dynamics of this potassium channel isoform, which can be accurately captured by a single gating variable. The activation gate is represented by a dimensionless variable \( m(t) \in [0, 1] \), denoting the fraction of activated potassium channels at time \( t \). Its dynamics follow a voltage-dependent relaxation toward a steady-state value:
\begin{equation}
\frac{dm}{dt} = \frac{m_{\infty}(V_m) - m}{\tau_m(V_m)},
\end{equation}
where \( m_{\infty}(V_m) \) is the steady-state activation curve, and \( \tau_m(V_m) \) is the voltage-dependent activation time constant. These functions are derived from experimental fits to  K\textsubscript{V}3.1 channel data and are given by:
\begin{align}
m_{\infty}(V_m) &= \frac{1}{1 + \exp\left(\frac{18.7 - V_m}{9.7}\right)}, \\
\tau_m(V_m) &= \frac{1}{1 + \exp\left(-\frac{V_m + 46.56}{44.14}\right)}.
\end{align}

The effective potassium conductance is proportional to the activation variable: $G_K = \bar{g}_K m,$ where \( \bar{g}_K \) is the maximal potassium conductance. The resulting potassium current is then computed using Ohm’s law as:
\begin{equation}
I_K = \bar{g}_K m (V_m - E_K),
\end{equation}
where \( E_K \) denotes the potassium reversal potential. This formulation captures the delayed rectification behavior of potassium channels, which contributes to repolarization and termination of the action potential following sodium channel activation.

Collecting all components, the full system of equations describing the hybrid neuron model is:

\begin{subequations}
\label{eq:complete-model}
\begin{align}
    \label{eq:dynamical-system}
    \frac{dV_m}{dt} &= \frac{1}{C_m} \left( I_{\text{ext}} - \bar{g}_{\text{Na}} (O_1 + O_2)(V_m - E_{\text{Na}}) - \bar{g}_K m (V_m - E_K) \right), \\
    \frac{dm}{dt} &= \frac{m_{\infty}(V_m) - m}{\tau_m(V_m)}, \\
    \frac{d\vec{s}}{dt} &= Q(V_m) \vec{s}.
\end{align}
\end{subequations}

The state space thus consists of  $(V_m, m, s_1, ..., s_6) \in \mathbb{R}^8$. To analyze stability, we find equilibrium points where $\frac{dV_m}{dt} = \frac{dm}{dt} = \frac{d\vec{s}}{dt} = 0$, linearize the system around these points and evaluate the stability of solutions by evaluating the eigenvalues of the Jacobean matrix. All of these steps are described in detail in the methods section.

\section{Methods}
\label{sec:methods}
To investigate the stability of the neuronal dynamics described by the coupled system of equations governing the membrane voltage, gating variable, and Markov-state probabilities, we perform a local stability analysis centered on the system’s fixed points. Specifically, we analyze the system defined by equations \ref{eq:complete-model}, which together define a nonlinear dynamical system in the variables \( V_m \), \( m \), and the six-state occupation density vector \( \vec{s} = [C_1, C_2, O_1, O_2, I_1, I_2]^T \). 

Fixed points \((V_m^*, m^*, \vec{s^*})\) of the system correspond to steady-state solutions where all time derivatives vanish. To numerically compute these fixed points, we employ a Newton–Raphson root-finding method applied to the membrane voltage derivative \( \frac{dV_m}{dt} \), solving for values of \( V_m^* \) such that the net current is zero. The algorithm is initialized with multiple guesses for \( V_m \) to ensure robust convergence. At each iteration, we first determine the steady state values of  the value of $m, \vec{s}$. $m$ is determined simply from equation \ref{eq:dynamical-system} as $m^* = m_{\infty}$. To find the steady state of the Markov model occupation probabilities, we need to solve the steady state equation:  
$
Q \vec{s} = \vec{0}.
$

But, since \(Q_{ii} = -\sum_{j \ne i} Q_{ji}\), $Q$ is rank deficient and hence non-invertible, as we can see by writing it down:

\small{
\[
\hspace{-2.5cm}
Q=
\begin{bmatrix}
-(r_{\mathrm{C}_2\mathrm{C}_1}+r_{\mathrm{I}_1\mathrm{C}_1}) & r_{\mathrm{C}_1\mathrm{C}_2} & 0 & 0 & r_{\mathrm{C}_1\mathrm{I}_1} & 0 \\[4pt]
r_{\mathrm{C}_2\mathrm{C}_1} & -(r_{\mathrm{C}_1\mathrm{C}_2}+r_{\mathrm{O}_1\mathrm{C}_2}+r_{\mathrm{O}_2\mathrm{C}_2}) & r_{\mathrm{C}_2\mathrm{O}_1} & r_{\mathrm{C}_2\mathrm{O}_2} & 0 & 0 \\[4pt]
0 & r_{\mathrm{O}_1\mathrm{C}_2} & -(r_{\mathrm{C}_2\mathrm{O}_1}+r_{\mathrm{I}_1\mathrm{O}_1}) & 0 & r_{\mathrm{O}_1\mathrm{I}_1} & 0 \\[4pt]
0 & r_{\mathrm{O}_2\mathrm{C}_2} & 0 & -r_{\mathrm{C}_2\mathrm{O}_2} & 0 & 0 \\[4pt]
r_{\mathrm{I}_1\mathrm{C}_1} & 0 & r_{\mathrm{I}_1\mathrm{O}_1} & 0 & -(r_{\mathrm{C}_1\mathrm{I}_1}+r_{\mathrm{O}_1\mathrm{I}_1}+r_{\mathrm{I}_2\mathrm{I}_1}) & r_{\mathrm{I}_1\mathrm{I}_2} \\[4pt]
0 & 0 & 0 & 0 & r_{\mathrm{I}_2\mathrm{I}_1} & -r_{\mathrm{I}_1\mathrm{I}_2}
\end{bmatrix}.
\]}
To remedy this deficiency and find the steady state occupation vector $\vec{s}$, we construct an modified transition matrix $A$, by replacing the equation of the transition rate of state $I_2$, by the constraint $\sum_{i=1}^6 s_i(t) = 1.$ This gives us the modified transition matrix $A$ as: 
$$
\resizebox{\textwidth}{!}{$
A = 
\begin{bmatrix}
-(C1C2 + C1I1) & C2C1           & 0                 & 0         & I1C1                & 0      \\
C1C2           & -(C2C1 + C2O1 + C2O2) & O1C2      & O2C2     & 0                   & 0      \\
0              & C2O1           & -(O1C2 + O1I1)     & 0         & I1O1                & 0      \\
0              & C2O2           & 0                 & -O2C2     & 0                   & 0      \\
C1I1           & 0              & O1I1              & 0         & -(I1C1 + I1O1 + I1I2) & I2I1   \\
1              & 1              & 1                 & 1         & 1                   & 1
\end{bmatrix},
$}
$$
and the matrix equation we use to find steady state occupation vector now becomes:
$$
A \vec{s} = \begin{bmatrix}
0\\ 0\\ 0\\ 0\\ 0\\ 1\\
\end{bmatrix}.
$$
We, then, find the steady state occupation fraction vector $\vec{s}$ as: 
$$
\vec{s^*} = A^{-1} \begin{bmatrix}
0\\ 0\\ 0\\ 0\\ 0\\ 1\\
\end{bmatrix}.
$$

Finally, using $m^*$ and $\vec{s^*}$  we just calculated, \( V_m \) is updated using the standard Newton update rule, where the first derivative is approximated via central scheme numerical differentiation. Convergence is declared once the absolute magnitude of \( \frac{dV_m}{dt} \) falls below a predefined tolerance (typically \( 10^{-6} \)). This method provides a reliable means to identify steady-state voltage solutions under various channel configurations and stimulation regimes.

To assess the local stability of each equilibrium point, we linearize the system around the fixed point \((V_m^*, m^*, \vec{s}^*)\) by computing the Jacobian matrix \( \mathbf{J} \), which captures how small perturbations to the system evolve over time. The Jacobian is constructed by evaluating the partial derivatives of the system's right-hand side with respect to each dynamic variable. 

All partial derivatives, including those of the Markov transition matrix \( \mathbf{Q}(V_m) \), the potassium gating functions \( m_\infty(V_m) \) and \( \tau_m(V_m) \), and the voltage-dependent currents, are computed numerically using a central difference approximation scheme and automatic differentiation.

Derivatives are computed for each dynamical variable independently. Then all derivative components are concatenated to form the full \(8 \times 8\) Jacobian matrix \( \mathbf{J} \), which is then evaluated at the fixed point. Eigenvalues of \( \mathbf{J} \) are used to determine local stability and detect bifurcations.

\begin{equation}
\resizebox{0.9\textwidth}{!}{$
J = 
\left[
\begin{array}{ccc}
\displaystyle -\frac{\bar{g}_{\mathrm{Na}} (O_1 + O_2) + g_K m}{c} 
& \displaystyle -\frac{\bar{g}_K (V_m - E_K)}{c}
& \begin{matrix} 0 & 0 & \displaystyle -\frac{\bar{g}_{\mathrm{Na}} (V_m - E_{\mathrm{Na}})}{c} & \displaystyle -\frac{\bar{g}_{\mathrm{Na}} (V_m - E_{\mathrm{Na}})}{c} & 0 & 0 \end{matrix} \\[2ex]

\displaystyle \frac{m'_{\infty} \tau_m - \tau_m' (m_\infty - m)}{\tau_m^2}
& \displaystyle -\frac{1}{\tau_m}
& \bm{0}_{1 \times 6} \\[2ex]

\frac{\partial \mathbf{Q}}{\partial V} \bm{S} 
& \bm{0}_{6 \times 1}
& \mathbf{Q}
\end{array}
\right]
$}
\end{equation}

By numerically computing time series solutions, using Euler scheme, of the system at different equilibrium points and for different parameter values, this system has been found to exhibit oscillatory solutions if at least one of the eigenvalues \( \lambda_i \) of the Jacobian matrix \( \mathbf{J} \) satisfy the condition \( \text{Re}(\lambda_i) \ge 0 \). Otherwise the system has a locally stable fixed point. These numerical experiments used  the $Na\textsubscript{V}1.1$ sodium isoform, choosing a window of 50 ms for the numerical simulation. The system has been deemed oscillatory if, once $V_m$ starts oscillating, it keeps oscillating for the rest of the simulation period. These oscillation were characterized by the existence of $V_m$ peak, i.e. a value greater than the preceding and following values in the time-series solution. This oscillatory vs. stationary behavior can be seen in figure \ref{fig:time-series}.

To explore the effect of biophysical parameters on neuronal dynamics, we perform a systematic sweep over the maximal sodium and potassium conductances \( \bar{g}_{\mathrm{Na}} \) and \( \bar{g}_{\mathrm{K}} \). These parameters are varied from 0 to 10 times of their physiological values. At each grid point in the parameter space, we compute the Jacobian and assess stability via its eigenvalues. We increase the stimulation current $I_{ext}$ at each point in the parameter space from $0$ mA/cm$^2$ till $50$ mA/cm$^2$ by increments of 1 mA/cm$^2$. We then record the minimal stimulation current needed to obtain an oscillatory solution.

To validate the predicted dynamical behavior for all sodium channel isoforms, we numerically integrated the full system of differential equations using a fourth-order Runge--Kutta method. Time-domain simulations were conducted at representative points in parameter space for each one of the isoforms, including regions predicted to exhibit stable equilibria or oscillatory limit cycles. The resulting trajectories confirmed the presence of both stationary and periodic solutions in their expected domains. 

Bifurcation boundaries in the parameter space are identified by systematically scanning across combinations of model parameters (i.e. \( (\bar{g}_{\text{Na}}, \bar{g}_{\text{K}}) \)) and monitoring changes in the qualitative behavior of the system’s fixed points. A bifurcation point is detected when the real part of an eigenvalue crosses zero, indicating a transition between a stable and an unstable solution. In particular, Hopf bifurcation lines are located by identifying parameter sets at which a pair of complex-conjugate eigenvalues crosses the imaginary axis, signifying the onset or disappearance of sustained oscillations.

To visualize the stability structure of the system across parameter space, we generated heatmaps over the \( (\bar{g}_{\mathrm{Na}}, \bar{g}_{\mathrm{K}}) \) plane. Each point on the grid corresponds to a specific combination of sodium and potassium maximal conductances, color-coded by the minimal external current \( I_{\mathrm{ext}} \) required to induce sustained oscillations. These heatmaps provide a clear depiction of the boundaries between regions of stationary and oscillatory dynamics, allowing the visualization of bifurcation structure and excitability landscapes in the model. 

All of the aforementioned computations are carried out using \texttt{Python}. Eigenvalues are computed using standard numerical linear algebra routines via the \texttt{numpy.linalg.eig} function, available in \texttt{numpy} package.

\section{Results}
\label{sec:results}

\begin{figure}[htb]
    \centering
    \includegraphics[width=0.7\linewidth]{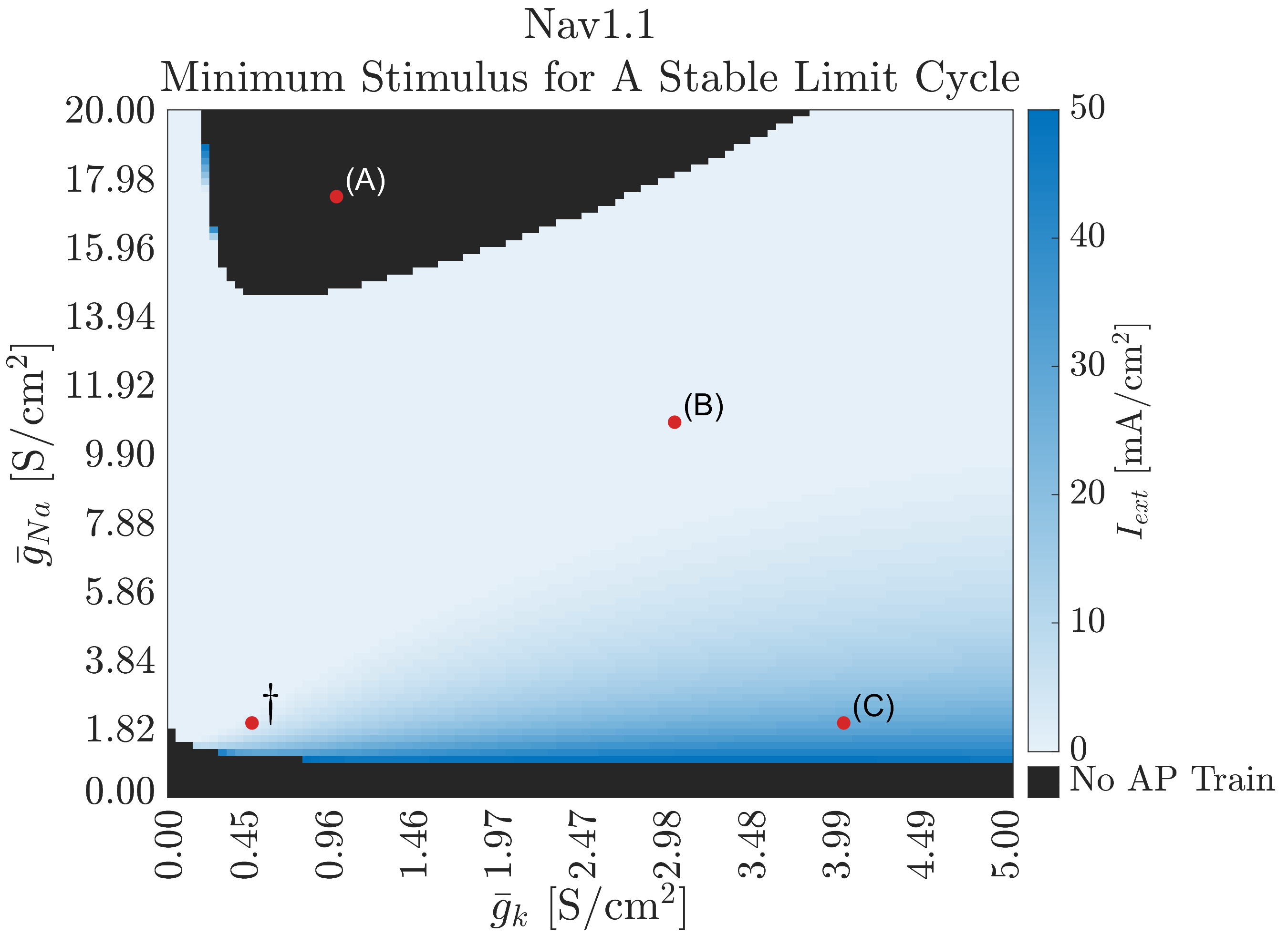}
    \caption{Minimum external stimulus current \( I_{\mathrm{ext}} \) required to induce stable oscillations across the parameter space defined by maximal sodium and potassium conductances \( (\bar{g}_{\mathrm{Na}}, \bar{g}_{\mathrm{K}}) \) for the Na\textsubscript{V}1.1– K\textsubscript{V}3.1 channel combination. Color intensity indicates the threshold stimulus required for a stable limit cycle, with darker regions indicating higher thresholds or stationary regimes. Black regions denote parameter combinations where no stable oscillations are observed for any tested \( I_{\mathrm{ext}} \). Red dots mark representative points A, B, and C used for time-domain validation (see Figure~\ref{fig:time-series}). The $\dagger$ point is sodium and potassium conductance in a typical neuron of a human that uses these specific potassium and sodium channel isoforms.}
    \label{fig:stability-heatmap}
\end{figure}

\begin{figure}
    \centering
    \includegraphics[width=1\linewidth]{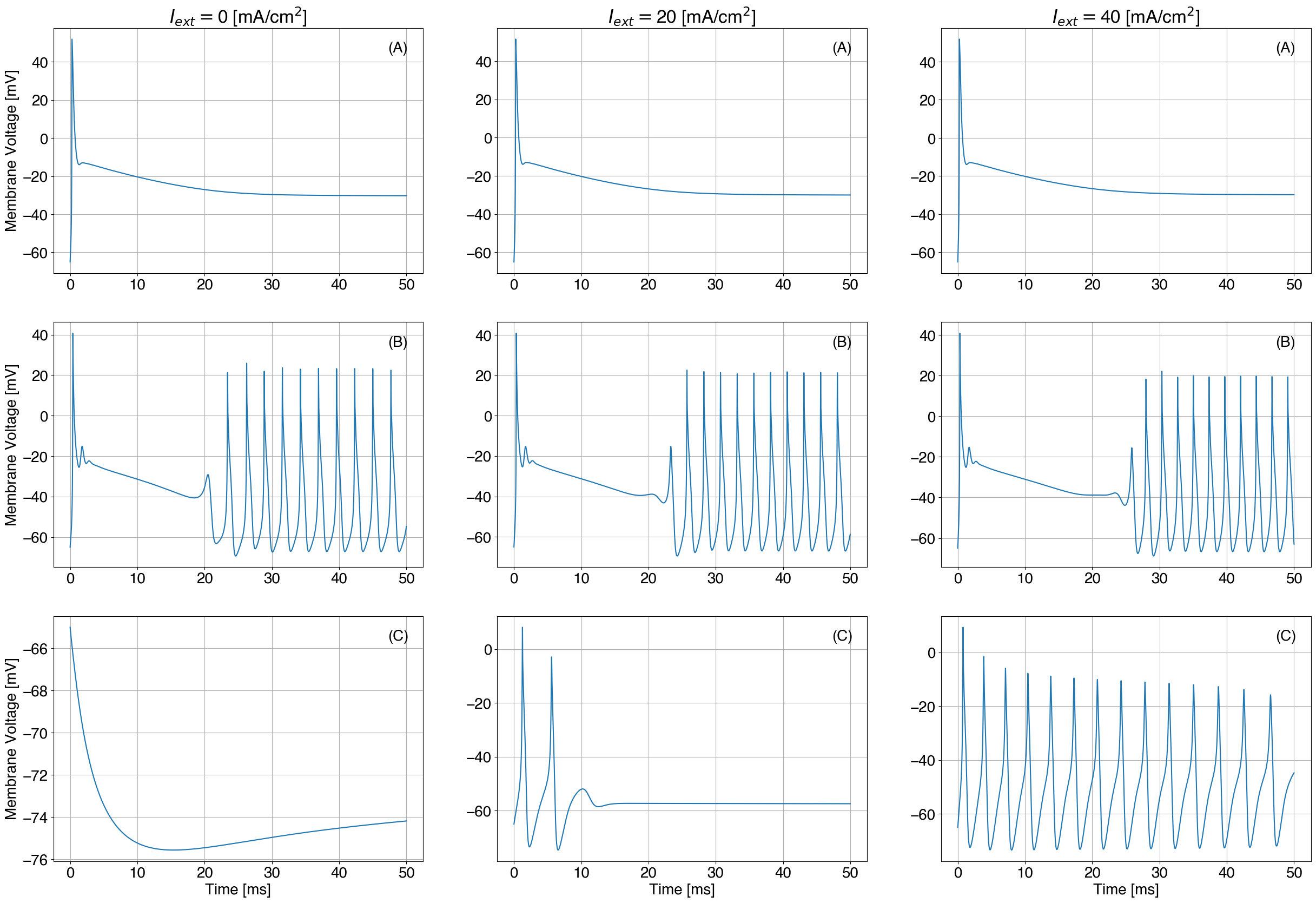}
    \caption{Time series simulations of membrane potential \( V_m(t) \) at three representative points (A, B, and C) in the conductance parameter space. The behavior of each point (oscillatory or stationary) confirms the predictions from eigenvalue-based stability analysis.}
    \label{fig:time-series}
\end{figure}

We investigated the dynamical behavior of the system across a range of sodium and potassium maximal conductances \((\bar{g}_{\mathrm{Na}}, \bar{g}_{\mathrm{K}})\) by computing the minimal external current \( I_{\mathrm{ext}} \) required to sustain stable oscillations at each parameter point, as mentioned in the methods section. The resulting bifurcation structure was visualized using heatmaps. Figure~\ref{fig:stability-heatmap} shows the excitability landscape for the Na\textsubscript{V}1.1 sodium channel isoform in combination with the K\textsubscript{V}3.1 potassium channel model. Brighter regions correspond to lower stimulation thresholds and higher excitability, while darker regions reflect parameter combinations that only yield stationary fixed points regardless of stimulation strength.
To confirm the predictions of the stability analysis, we performed time-domain simulations at selected points (labeled A, B, and C) on the heatmap. Figure~\ref{fig:time-series} shows the resulting membrane potential traces: Point A remains quiescent across all inputs, Point B exhibits robust oscillations at all tested stimulation levels, while Point C transitions from damped subthreshold responses to stable oscillations only when \( I_{\mathrm{ext}} > 30\,\text{mA/cm$^2$} \).

\begin{figure}[h!]
    \centering
    \includegraphics[width=0.75\linewidth]{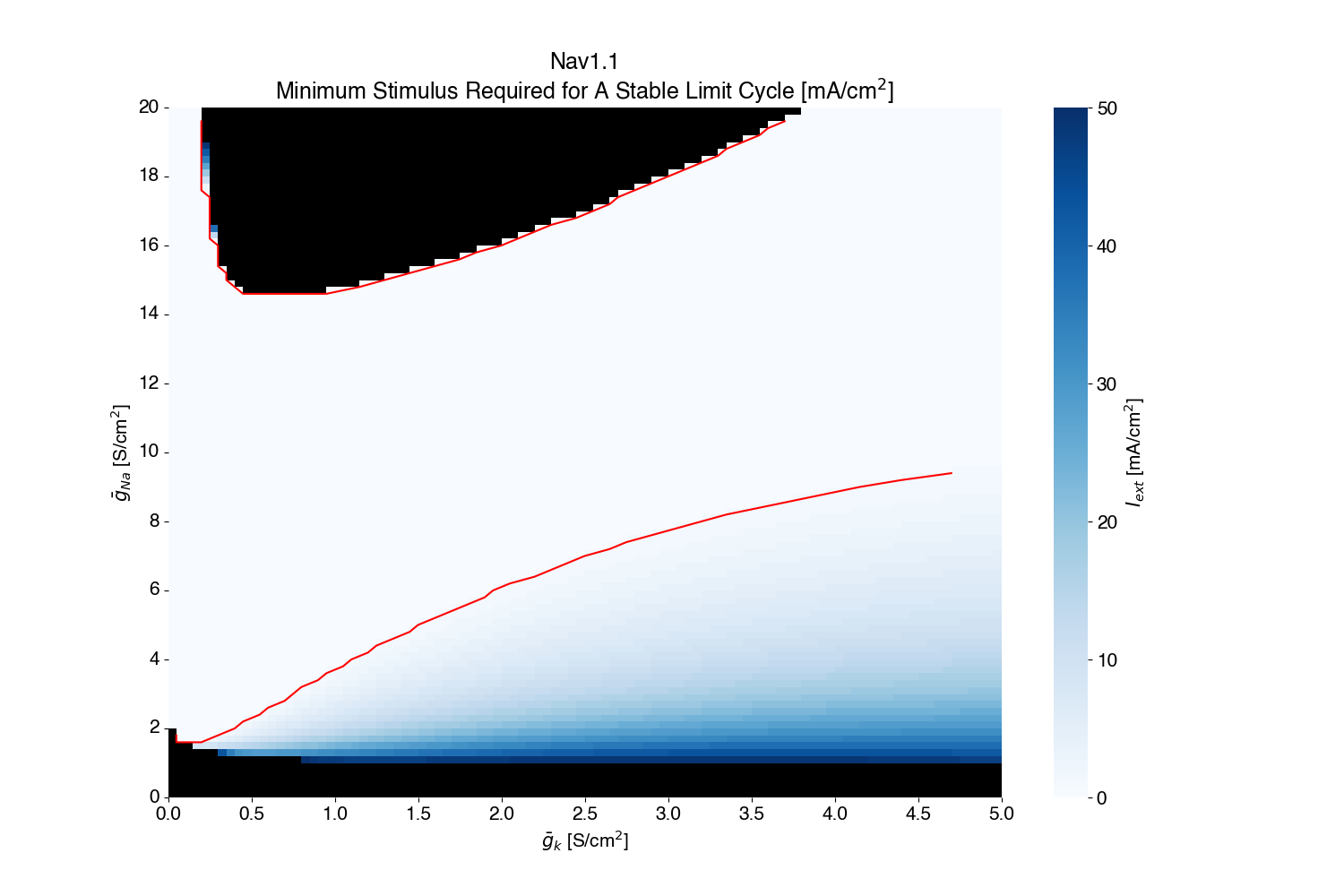}
    \caption{Bifurcation structure of the Markov model of VGSC isoform Na\textsubscript{V}1.1 in the \( (\bar{g}_{\mathrm{K}}, \bar{g}_{\mathrm{Na}}) \) conductivity space. The heatmap shows the minimum external current \( I_{\mathrm{ext}} \) [nA] required to sustain a stable limit cycle. The red curve marks the Hopf bifurcation boundary separating excitable (oscillatory) and quiescent (stationary) regimes.}
    \label{fig:hh-bifurcation}
\end{figure}

Applying bifurcation analysis to the hybrid model comprising the Na\textsubscript{V}1.1 Sodium Markov Model scheme and the K\textsubscript{V}3.1 Hodgkin–Huxley potassium channel yields the heatmap in Fig.~\ref{fig:hh-bifurcation}.  This figure illustrates the dynamical structure of the system across the two‑dimensional parameter space spanned by the maximal sodium (\(\bar{g}_{\mathrm{Na}}\)) and potassium (\(\bar{g}_{\mathrm{K}}\)) conductances.  Each point is colored by the minimal external current \(I_{\mathrm{ext}}\) needed to sustain a stable limit cycle, with darker shades indicating higher thresholds or absence of oscillations.  A red curve superimposed on the heatmap marks the Hopf bifurcation boundary separating spontaneous and stimulated oscillation regimes.

\begin{figure}[h!]
    \centering
    \includegraphics[width=1\linewidth]{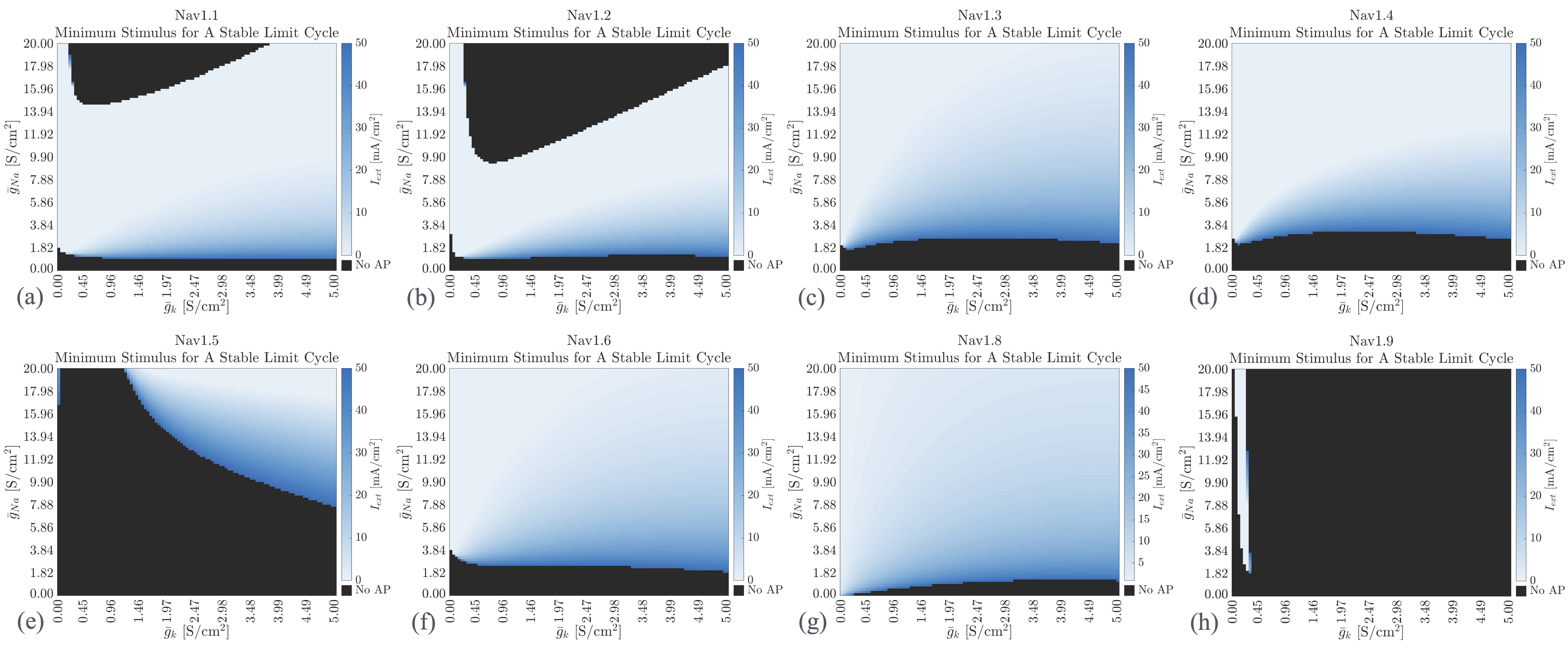}
    \caption{Excitability heatmaps of  (a) Na\textsubscript{V}1.1, (b) Na\textsubscript{V}1.2, (c) Na\textsubscript{V}1.3, (d) Na\textsubscript{V}1.4, (e) Na\textsubscript{V}1.5, (f) Na\textsubscript{V}1.6, (g) Na\textsubscript{V}1.8, (h) Na\textsubscript{V}1.9
    showing minimum excitation current needed to sustain a train of action potentials at different \(\bar{g}_{\text{Na}}\) and \(\bar{g}_K\) parameter value combinations.}
    \label{fig:nav}
\end{figure}

To systematically assess how different sodium channel isoforms modulate excitability, we generated excitability heatmaps for eight Na\textsubscript{V} subtypes—Na\textsubscript{V}1.1 through Na\textsubscript{V}1.6, along with Na\textsubscript{V}1.8 and Na\textsubscript{V}1.9—each paired with the same K\textsubscript{V}3.1 potassium channel model (Fig.~\ref{fig:nav}).  The isoform Na\textsubscript{V}1.7 is not plotted because its heatmap is completely black, indicating no oscillatory behavior in the tested parameter range.  Each panel in Fig.~\ref{fig:nav} shows the minimal external current \(I_{\mathrm{ext}}\) required to induce sustained oscillatory activity across a grid of maximal sodium and potassium conductance values.  These heat maps reveal substantial variability in excitability profiles across isoforms: some, like Na\textsubscript{V}1.6, display broad regions of low-threshold oscillations (Fig.~\ref{fig:nav}(f)), while some others like Na\textsubscript{V}1.7 (not shown) fail to support any oscillatory behavior within the tested parameter range.  Na\textsubscript{V}1.1 and Na\textsubscript{V}1.2 (Figs.~\ref{fig:nav}(a) and (b)) are both expressed in the CNS and show similar inactivity regions at high values of maximal sodium conductance.  The other isoforms expressed in the CNS—Na\textsubscript{V}1.3 and Na\textsubscript{V}1.6 (Figs.~\ref{fig:nav}(c) and (f))—as well as Na\textsubscript{V}1.4 (Fig.~\ref{fig:nav}(d)), which is expressed in skeletal muscle, and Na\textsubscript{V}1.8 (Fig.~\ref{fig:nav}(g)), which is expressed in the PNS, exhibit very low thresholds for stable oscillations.  Finally, Na\textsubscript{V}1.5 (Fig.~\ref{fig:nav}(e)), expressed in cardiac muscle, and Na\textsubscript{V}1.9 (Fig.~\ref{fig:nav}(h)), which is mostly restricted to the PNS, show very low activity and do not possess stable oscillatory solutions in most of the parameter space \cite{catterall2005international, goldin1999diversity}.

\section{Conclusion}
\label{sec:conclusion}

Our analysis demonstrates that the excitability of a conductance-based neuronal model is influenced by the specific voltage-gated sodium channel (VGSC) isoform employed, in conjunction with a fixed voltage-gated potassium channel (VGPC). By systematically varying the maximal conductances $\bar{g}_{\mathrm{Na}}$ and $\bar{g}_{\mathrm{K}}$ across physiologically plausible ranges, we constructed excitability heatmaps for nine human VGSC isoforms paired with the fast-activating K\textsubscript{V}3.1 potassium channel. These maps delineate the minimal external stimulus current $I_{\mathrm{ext}}$ required to initiate sustained oscillatory activity, revealing marked isoform-specific differences in excitability profiles.

The results clearly show that not all VGSC isoforms support repetitive firing across the same region of conductance parameter space. Certain isoforms, such as Na\textsubscript{V}1.3 and Na\textsubscript{V}1.6 which are mostly expressed in the CNS and Na\textsubscript{V}1.4 which is expressed in skeletal muscles, exhibit robust excitability over a wide range of sodium and potassium conductance values, requiring relatively low external current to elicit action potentials (Figs. \ref{fig:nav}(c), (f), and (d) respectively). This suggests a high degree of reliability and dynamic range, making these isoforms attractive candidates for applications in synthetic neuron design where repetitive firing with low stimulus is required. Conversely, isoforms expressed in the PNS such as Na\textsubscript{V}1.7 shows a complete lack of oscillatory behavior within the investigated parameter range, and Na\textsubscript{V}1.9 supports oscillatory activity in a very narrow range of parameters  (Fig. \ref{fig:nav}(h)). This implies that their kinetic properties constrain the generation of action potentials under typical physiological conditions, possibly reflecting a specialization for low activity roles or narrowly tuned excitability in vivo. The latter two isoforms as well as Na\textsubscript{V}1.5 (Figs.\ref{fig:nav}(e)), which is expressed in cardiac muscle, can be good candidates in contexts requiring higher stimulus for firing or few action potentials rather than a continuous train of action potentials at the desired stimulation level. 

A key contribution of this work is the systematic comparison of excitability landscapes across voltage‑gated sodium channel isoforms within a unified modeling and bifurcation‑analysis framework. This approach yields a mechanistic interpretation of how isoform‑specific kinetics shape neuronal firing behavior and provides practical guidance for experimental design: Na\textsubscript{V} isoforms that support broad excitable regimes with low activation thresholds are attractive candidates for synthetic neuron fabrication, particularly when conductance values cannot be precisely tuned.

Our current analysis is restricted to a single potassium channel type, K\textsubscript{V}3.1, which facilitates controlled comparisons but does not encompass the rich diversity of K\textsubscript{V} channel kinetics in biological neurons. Although employing a Hodgkin–Huxley–type model for the potassium channel alongside a Markov model for Na\textsubscript{V} isoforms preserves the conceptual integrity of the hybrid model since HH models arise as limiting cases of Markov models when certain transition rates become large, this simplification excludes the multiple activation and inactivation dynamics of other potassium isoforms. Extending the framework to test Na\textsubscript{V} isoforms in combination with alternative K\textsubscript{V} isoforms (e.g., K\textsubscript{V}1.2 or K\textsubscript{V}4.x) would provide a more comprehensive picture of excitability. The model also omits synaptic variability and neuronal morphology, and the bifurcation analysis focuses on qualitative transitions in excitability rather than quantitative measures such as firing rate or spike amplitude, limiting the applicability of the results to more complex physiological settings. Nonetheless, our findings establish a systematic framework for evaluating the influence of voltage‑gated sodium channel isoform identity on neuronal excitability. The bifurcation heatmaps serve as functional guides to sodium–potassium channel combinations and offer a practical tool for designing synthetic excitable systems. Future work may incorporate a broader range of ion channels, temperature effects, isoform‑specific mutations, and morphological complexity, and may extend the framework to networks of neurons to explore the conditions that give rise to specific collective dynamics.



 \section*{Acknowledgments}
We would like to thank Dr. Francois St-Pierre, Dr. Matthew Shorey, Dr. Allen Liu, Dr. Nathan Cahill, and Dr. Laura Munoz for useful discussions and feedback.
This work was supported by NSF Award 1935277.

\bibliographystyle{siamplain}
\bibliography{references}

\end{document}

%% file: ex_shared.tex
\usepackage{lipsum}
\usepackage{amsfonts}
\usepackage{graphicx}
\usepackage{epstopdf}
\usepackage{algorithmic}
\ifpdf
  \DeclareGraphicsExtensions{.eps,.pdf,.png,.jpg}
\else
  \DeclareGraphicsExtensions{.eps}
\fi

\usepackage{enumitem}
\setlist[enumerate]{leftmargin=.5in}
\setlist[itemize]{leftmargin=.5in}

\newsiamremark{remark}{Remark}
\newsiamremark{hypothesis}{Hypothesis}
\crefname{hypothesis}{Hypothesis}{Hypotheses}
\newsiamthm{claim}{Claim}
\newsiamremark{fact}{Fact}
\crefname{fact}{Fact}{Facts}

\headers{A Stability Analysis of Action Potential Generation}{Y. Abdullah, V. Hart and M. Das}

\title{Stability analysis of action potential generation using Markov models of voltage‑gated sodium channel isoforms}

\author{Youssof Abdullah\thanks{School of Mathematics and Statistics, Rochester Institute of Technology 
  (\email{ya1753@rit.edu})}
\and Violet Hart\thanks{School of Physics, Rochester Institute of Technology.}
\and Moumita Das\thanks{School of Physics and Astronomy \& School of Mathematics and Statistics, Rochester Institute of Technology. (\email{modsps@rit.edu})}}

\usepackage{amsopn}